\documentclass{article}
\usepackage[colorinlistoftodos]{todonotes}

\usepackage{amsmath,graphicx}
\usepackage{graphicx}			
\usepackage[utf8]{inputenc}
\usepackage{psfrag,epsfig,graphics}
\usepackage{amsmath,amsthm,amssymb,multirow}
\usepackage{mathbbol}
\usepackage{amssymb}             

\usepackage{array}
\usepackage{mathabx} 
\usepackage{tikz, pgfplots}

\DeclareSymbolFontAlphabet{\amsmathbb}{AMSb}%

\usepackage{graphicx}

\usepackage[noadjust]{cite}
\usepackage{multirow}
\usepackage{hyperref}

\usepackage[lined,linesnumbered,ruled]{algorithm2e}

\usepackage{color}  

\newcommand{\bs}{\boldsymbol}

\newcommand{\vx}{\bs{x}}

\title{A Pathology-Based Machine Learning Method to Assist in Epithelial Dysplasia Diagnosis}

\author{Karoline da Rocha\footnotemark[4]\thanks{Corresponding author. E-mail address: karolrocha0802@gmail.com (K. Rocha) }, Jos\'e C. M. Bermudez\footnotemark[4], Elena R. C. Rivero\footnotemark[3], M\'arcio H. Costa\footnotemark[4]
	\\ \\ \footnotemark[4]{\, \footnotesize{Graduate Program in Electrical Engineering, Federal University of Santa Catarina, Florian\'opolis, Brazil}}
	\\ \footnotemark[3]{\footnotesize{\, Department of Pathology, Federal University of Santa Catarina, Florian\'opolis, Brazil}}
	}%

\begin{document}
%
\maketitle
\begin{abstract}
The Epithelial Dysplasia (ED) is a tissue alteration commonly present in lesions preceding oral cancer, being its presence one of the most important factors in the progression toward carcinoma. This study proposes a method to design a low computational cost classification system to support the detection of dysplastic epithelia, contributing to reduce the variability of pathologist assessments. We employ a multilayer artificial neural network (MLP-ANN) and defining the regions of the epithelium to be assessed based on the knowledge of the pathologist. The performance of the proposed solution was statistically evaluated. The implemented MLP-ANN presented an average accuracy of $87\%$, with a variability much inferior to that obtained from three trained evaluators. Moreover, the proposed solution led to results which are very close to those obtained using a convolutional neural network (CNN) implemented by transfer learning, with 100 times less computational complexity. In conclusion, our results show that a simple neural network structure can lead to a performance equivalent to that of much more complex structures, which are routinely used in the literature.

\vspace{3ex}

\end{abstract}
%
%
\section{Introduction}
\label{sec:intro}

\hspace{\parindent}The Squamous cell carcinoma (SCC) is one of the most common oral cavity cancers among men. Survival rates for this cancer are low \cite{Nag2018_Analysis,Boockwho}. Thus, early diagnosis is essential for survival and for determining the most appropriate treatment. Epithelial Dysplasia (ED) is a very frequent tissue alteration in lesions that precede oral cancer. Its cause may be associated with etiological factors, mainly with the consumption of tabacco and alcohol \cite{introd1}.

The standard ED diagnosis is made by pathologists through analysis of histopathological slides in search of changes in the epithelium. These alterations may be architectural, such as basilar hyperplasia, droplet-shaped epithelial projections, increased number of mitoses or loss of cellular cohesion. They may also be cellular changes, such as enlarged nuclei, enlarged cells, increased nuclear-cytoplasmic ratio, or atypical mitotic figures \cite{DE_conceito}.

The most widely employed criteria for grading oral ED are those defined in the World Health Organization (WHO) classification system, which considers the presence of certain architectural and cytological features. The more prominent and numerous these features are identified in the histopathological image, the more severe the grade of dysplasia~\cite{Warnakulasuriya2008_Oral,Boockwho}. However, the assessment of dysplasia is subjective and strongly dependent on the personal experience of the pathologist~\cite{embalo2021evaluation}, and the diagnosis variability is well documented~\cite{Bouquot2006_Epithelial,Kujan2006_Evaluation,Warnakulasuriya2008_Oral,mahmood2021artificial}.

The area of computer-aided diagnosis has been growing over the past decades. Histopathological images are analyzed using quantitative measures of image characteristics. These measures are employed either by image processing algorithms or interpreted by machine learning systems to yield a suggested diagnosis. Such automated diagnostic tools can be an important aid to pathologists, enabling the creation of large-scale decision support systems to identify potentially malignant lesions~\cite{pallua2020future,mahmood2021artificial}. Some recent works have proposed automated diagnostic solutions to assist in the evaluation of histopathological images for the detection of epithelial dysplasia~\cite{artgo2,art1,art2,gupta2019tissue,Adel2018,Sami2009_ComputerAided,prabavathy2021analysis,baik2014automated,krishnan2009automated}.

In a previous work, texture characteristics of the epithelium have been used~\cite{artgo2} to classify histopathological images into normal epithelia and oral sub-mucous fibrosis (OSF) with or without dysplasia. In this article, textural characteristics of the epithelium were extracted using higher order spectra (HOS), local binary pattern (LBP), and laws texture energy (LTE). Five different classifiers were employed: Decision Tree (DT), Sugeno Fuzzy, Gaussian Mixture Model (GMM), K-Nearest Neighbor (K-NN), and Radial Basis Probabilistic Neural Network (RBPNN). The results of this work indicate that combination of texture and HOS features coupled with a fuzzy classifier resulted in $95.7\%$ accuracy. However, it is noted in the article that images need considerable processing to be used in the proposed classification method. In addition, the training and testing (evaluation) of the classifiers was performed using stratified 3-fold cross-validation, thus separating the data set into only three subsets, which may favor the occurrence of strong correlation between the training set and the test set. This evaluation approach leaves the performance results obtained ($95.7 \%$ accuracy) open to questioning.

The Block Intensity Code Comparison (BICC) was used in another study~\cite{art1} to extract the characteristics of the epithelium. Each image was divided into blocks and the intensity of each block was calculated. Blocks of sizes $5\times5$, $10\times10$ and $15\times15$ have been tested. The classifier used was a Radial Basis Function Neural Network (RBFNN) with a single hidden layer, and Gaussian activation function, classifying epithelia into normal or dysplastic. Different values were tested for the centers (means) of the Gaussians. These centers were located using k-Means, and $k=6$ led to the best result. The article fails to clearly explain how the final classification was made.
 
A convolutional neural network (CNN) has also been used~\cite{art2} to extract characteristics of histological images of the uterine cervix epithelium. The authors classify the epithelium in four classes: normal, mild dysplasia, moderate dysplasia and severe dysplasia. Each of the $66$ images available was subdivided into $10$ vertical segments. The images had to be correctly positioned so that each segment had the three layers of the epithelium. Each of these segments was subdivided into three parts of the same size (top, middle, bottom). Finally, $32\times32$ pixel patches were extracted from each part of each segment using non-overlapping windows, which generated a total of $75,000$ samples of size $32\times32$. Three CNNs were implemented, each of which classified the samples of each of the 3 parts into one of the 4 classes. These networks extract the characteristics of the samples and provide an initial classification.
The authors did not inform the amount of samples used to train and test the networks. The extracted characteristics fed 5 different classifiers (SVM, LDA, MLP, logistic regression and radom forest), which merged the characteristics of the 3 parts, obtained from the CNNs, and classified the entire epithelium image. The best performance ($77.25\%$ accuracy) was obtained with logistic regression and random forest, and using leave-one-out cross-validation performed only once to classify the 66 images.

Also using CNN, Gupta et al.~\cite{gupta2019tissue} classified epithelial images into 4 classes (normal, mild dysplasia, moderate dysplasia or severe dysplasia). The CNN received epithelial images as inputs. The classifier was trained using off-the-shelf packages. No details were given regarding the characteristics of the network nor the training algorithm. There were 2688 images taken from 672 tissue images of 52 patients, with approximately the same amount of images belonging to each of the four classes. The CNN training was performed from scratch using 70\% of the available samples. The remaining 30\% were used for testing. The CNN trained over 75 epochs presented an accuracy of $89.3\%$ for the test set.

Classification based on extracted values of the 16 WHO defined features was proposed in another study~\cite{Adel2018}. No detail was provided on how the extraction was performed. An SVM and a K-nearest-neighbor classifiers were tested using the different feature sets. The reported results were based on a single classification run using 46 images, from which 32 were used for training an 14 for test in the SVM implementation. The accuracy results varied from 71.4\% to 78.6\% in most implementations, and reached 92.8\% only for the SVM classifier operating on the features extracted by the Oriented FAST and Rotated BRIEF (ORB) algorithm.

Classification based on the similarity of neighboring rete ridges was proposed~\cite{Sami2009_ComputerAided}. Epithelium was classified as normal, dysplastic or carcinoma in-situ (CIS). The method was based on comparing the drop-shaped similarity level between the best matching pair of neighboring rete ridges. A contour extraction method was proposed, and the roundness of extracted twin contours was quantified. Clustering of the three classes was based on the roundness absolute values and differences.  Method illustration was based on a set of 17 images. No statistical evaluation was presented regarding the accuracy of the proposed classification method. 

In a recent publication, two characteristic extraction techniques were employed~\cite{prabavathy2021analysis},  namely histogram oriented gradient (HOG) and local binary pattern (LBP), were applied to discrete wavelet transformed $512\times512$ epithelial images. A three-layer back propagation neural network (BPNN) classified the images into normal or dysplastic epithelium. The best results were obtained using the HOG features, yielding an accuracy of $85\%$. The article does not detail the methodology used for the classification.

A semi-automatic algorithm has been also proposed~\cite{baik2014automated} to predict the progression of oral premalignant lesions (OPL) to invasive squamous cell carcinoma (SCC). The authors initially use two Random Forests, one to segment the nuclei of the histopathological images and the other to classify the nuclei found into normal and abnormal (cancerous). After the identification of the nuclei, a Nucleus Phenotype Score (NPS) was calculated based on the voting score that each nucleus received from the Random Forests classifier. Based on the average NPS of all image cores, an automated Tissue Nuclear Phenotype Score (aNPS) was assigned in order to identify OPL with high risk of progression. During the prediction of the progression of $71$ lesions in the test set, a $78\%$ sensitivity and a $71\%$ specificity were obtained.

Segmentation and classification of sub-epithelial connective tissue cells into normal or with oral submucous fibrosis (OSF) was proposed~\cite{krishnan2009automated}. Limiarization with several levels was used in the segmentation. An SVM-based classifier was used to classify the cells as per their geometric shape features (eccentricity and compactness). The classifier presented an accuracy of $88.69\%$ for the test set.

One common problem to these previous proposals is that they rely almost entirely on the information provided by the collected data. Despite the large popularity of machine learning (ML) algorithms, it is known that they tend to be data inefficient, and frequently generalize poorly to unseen cases. This characteristic is especially present in medical applications where the amount of available data is usually limited. Most ML algorithms that process raw data lead to classifications based on the correlation structure of the data presented to them during training. The success of this approach depends on a huge amount of data. Moreover, the diagnosis quality is also frequently dependent on factors independent of the data correlation, such as the amount and type of noise present in the data, the quality of data acquisition, and the amount of useful information embedded in the data. For instance, some of the presented solutions just discussed proposed to extract the characteristics of the epithelium by analyzing the entire image, which increases the possibility of using information that may hamper the classification process. This also tends to lead to a lack of consensus on the most relevant characteristics for classification. Another consideration usually made~\cite{Kujan2006_Evaluation,embalo2021evaluation} is that a binary classification system tends to be more helpful to the clinician for making critical clinical decisions in cases of high-risk epithelial dysplasia than the WHO three-level classification.

A more sensible approach is to complement the information provided by the raw data with the knowledge of experts in the application field. This tends to lead to more robust performances using simpler algorithms and less data. This work is a contribution in such a direction. We combine the knowledge of the pathologist with the information provided by the data to generate a method of detecting oral dysplasia which performs on a smaller amount of data, and is more robust to changes in the statistical characteristics of the data. We use the expert's knowledge to define the information to be delivered to an ML algorithm that aims at classifying histopathological images into dysplastic and non-dysplastic epithelia.




\section{Materials and methods}

\hspace{\parindent}In this section, we describe the database employed and the methods utilized.

\subsection{Database and sample preparation}

\hspace{\parindent}The database consists of histologic images obtained from the Biobank Archives of the Oral Pathology Laboratory of UFSC (CONEP B-$051$; Process No.~$25000.237810$/$2014$-$54$). The research project was approved by the UFSC Human Beings Research Ethics Committee (Platform Brasil under number CAAE $15025319.3.0000.0121$).

The images used were from histologic slides stained with hematoxylin and eosin. These slides were photographed with an original magnification of $200$x, using a digital camera coupled to an optical microscope. After being scanned, the images were duly anonymized to be used in this research.

The data set corresponded to $73$ cases, totaling $172$ images. Of these, $36$ cases ($88$ images) were from oral potentially malignant disorders with dysplastic epithelia and $37$ cases ($84$ images) were from fibrous hyperplasia with non-dysplastic epithelium.  

As evidenced in the Introduction, there seems to be no consensus yet regarding a good set of characteristics to be used for classification of dysplastic epithelia. Hence, we decided to use the image cutouts themselves as the information to be delivered to the classifier.

\begin{figure}[h]
\centering
\includegraphics[width=0.5\textwidth]{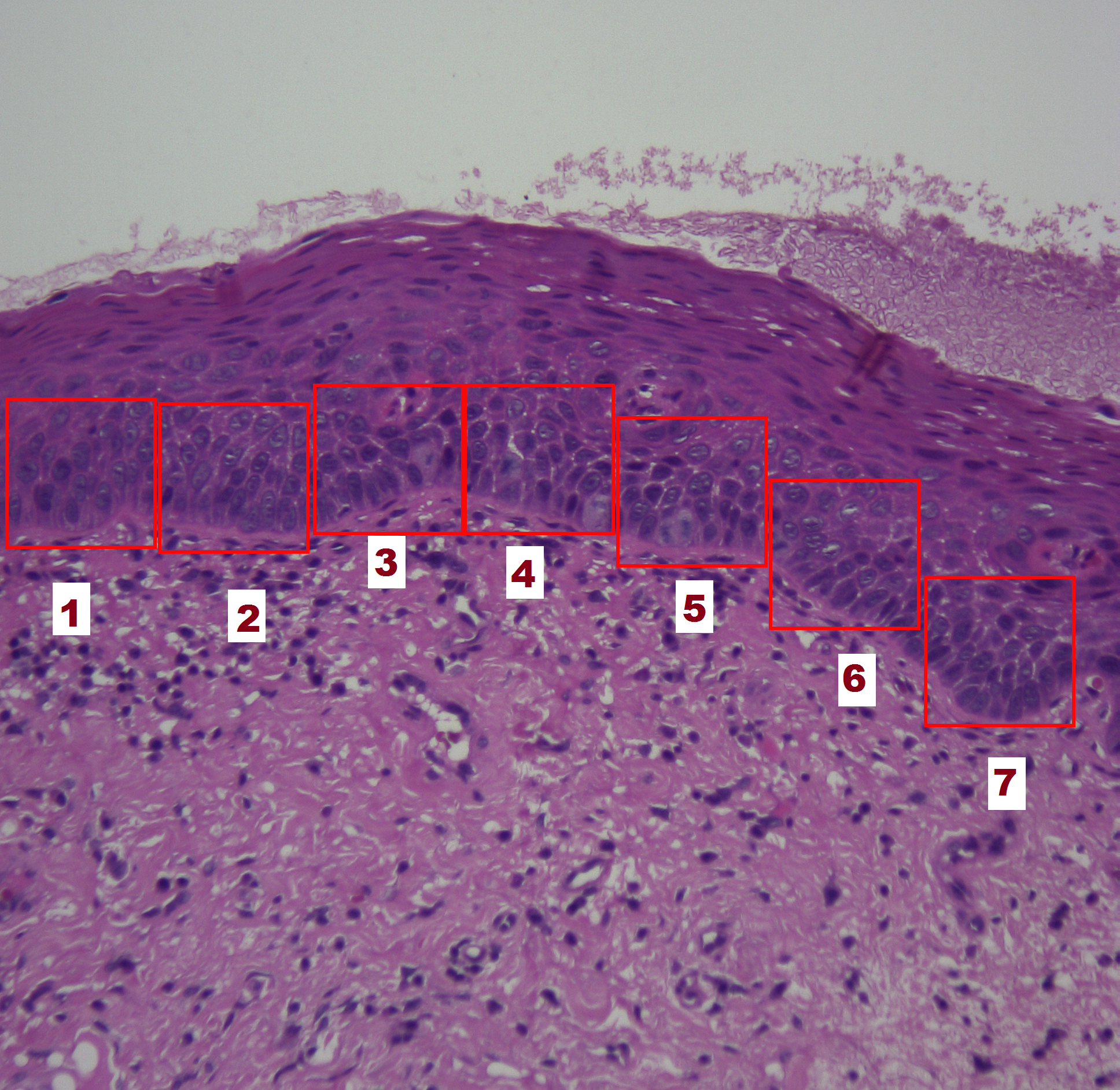}
\caption{Example of $7$ selected cutouts on the epithelium image.}
\label{img_regRecorte}
\end{figure}

To take advantage of the existing knowledge, we have chosen to extract cutouts from the image region containing the main visual characteristics used by the pathologist for such evaluations. The cutouts were extracted from regions located close to the border that separates the epithelial tissue from the connective tissue, as illustrated in Figure \ref{img_regRecorte}. The reasoning for this choice was that the lower third of a dysplastic epithelium image will always present detectable changes, regardless of the dysplasia degree (mild, moderate or severe). Through this informed choice, the use of a small image region simplifies the processing without harming the classification potential.

The definitions of location and size of each cutout were influenced by a combination of factors:
\begin{itemize}
	\item[a)] The use of small regions simplifies the data processing.
	\item[b)] The cutout image should contain enough information for a classification by the pathologist.
	\item[c)] A basic classifier should yield a reasonably low classification error using the cutouts. 
\end{itemize}

Based on these criteria, we have defined cutouts of arbitrary dimensions, at the frontier between the epithelial and connective tissues, with most of its area in the epithelial region, as shown in Figure~\ref{img_regRecorte}.

For processing by the classifier, each cutout image was converted from RGB to grayscale, reducing by $2/3$ the amount of data to be processed. Then, the cutout image values were normalized to increase robustness to intensity variations due to lighting effects in different acquisitions. Pixel intensities of each cutout were normalized to be in the range $[0,1]$. Finally, each cutout used in the classifier training stage was rotated three times by $90^\circ$, yielding four images for each cutout. Rotations by non-integer multiples of $90^\circ$ were not considered to avoid the need for further processing at the corners of the rotated cutouts. The use of four rotated positions was considered sufficient to provide the necessary robustness to image rotation.

\subsection{Methodologies}

\hspace{\parindent}This section describes the methods employed in the different steps of the classification task.

This project was implemented in Matlab$^\copyright$ (The MathWorks, Inc. Software). It was employed a personal computer with windows $10$ operating system, a $1.8$ GHz Intel Core $i5$ processor and $8$ GBytes of RAM memory.

\subsubsection{Definition of the classifier input}\label{sec:phase1}

\hspace{\parindent}The classifiers were implemented using feed-forward multi-layer neural networks. To define the structures of these networks it was necessary to choose, besides the size of the input cutouts, the number of hidden layers, the number of neurons in each hidden layer, the cost function to be used, and other necessary parameters depending on the previous choices. Different structures were considered and evaluated in the training stage. This stage was subdivided into two phases.

In the first phase we have implemented very simple networks with only one hidden layer. The objective of this phase was a first evaluation of the amount of information required for classification purposes, which would help to define the size of the input cutout.

The networks have been tested with $16,384$ and $65,536$ input neurons. Cutout dimensions were, respectively, $128\times128$ pixels and $256\times256$ pixels. In addition, the number of neurons in the hidden layer was varied in the set $\{20, 50, 100\}$. The cost function was the Mean Squared Error (MSE), given that the data had no outliers and the objective was an initial evaluation of the classification performance as a function of cutout size. The performance was evaluated based on the classification of each cutout individually, independent of the case to which they belonged. For the training, 1440 cutouts (720 with dysplasia and 720 without dysplasia) were randomly selected from the 1840 cutouts available. For testing, 160 randomly selected cutouts had been previously separated (being 80 with and 80 without dysplasia). From each of these sets of 80 cutouts, 50 were randomly selected. A total of 100 realizations, comprising training and test, were carried on to evaluate the performance. The results of the 100 realizations were averaged to determine the confusion matrix, specificity, sensitivity and average accuracy rate of the classifications. Based on these results, it was verified that the structures with $65,536$ input neurons yielded better performance on the average. Thus, all the structures implemented in the second phase had this same amount of neurons in the input layer.

The second phase of the design will be described next.

\subsubsection{Definition of the classifier structure}\label{sec:2phase}

\hspace{\parindent}The second design phase aimed at improving the performance of the classifier, given an input layer with $65,536$ neurons. We have initially used networks with two hidden layers. The number of neurons per layer was varied within the set $\{20, 50, 100, 150\}$. The cost functions tested were MSE and the cross-entropy. Another difference from the first phase was that the data unit considered was cases rather than individual cutouts. As we had data for 42 cases, at each realization the classifier was trained with the cutouts corresponding to 41 cases, and tested with the cutouts of the remaining case, thus using, leave-one-out cross validation. As the number of cutouts available was not exactly the same for all cases, care was taken to maintain the same number of dysplastic and non-dysplastic cutouts during training. The classification decision (dysplastic or non-dysplastic) was made by majority of the cutout classifications for each case. The confusion matrices, sensitivities, specificities, accuracy, and a novel figure of merit $D$ to be defined later were evaluated for each of the cases after test.

For statistical evaluation purposes, the complete process described above was repeated 50 times with randomly selected network initialization. Then, the confusion matrices, sensitivities, specificities, accuracy, and the figure of merit $D$ were averaged over 50 realizations for each of the 42 cases.

After evaluating the performance of the structures with two hidden layers, those which yielded better accuracy results had their number of hidden layers increased to verify if this increase would lead to an improvement of the average accuracy for the 42 cases under test. For each structure, the number of hidden layers was incremented until a drop in classification performance was verified. The training and test were performed exactly as done for the two-layer structures.

In both design phases, the classifiers were trained using the scaled conjugate gradient algorithm. The stop criterion was the, early stopping, in which the training was stopped when any of the following conditions occurred:

\begin{itemize}
	\item[a)] The maximum number of $1000$ iterations was reached.
	\item[b)] The error was equal to zero.
	\item[c)] The gradient was less than or equal to $10^{-6}$. 
	\item[d)] The validation set error increased in $6$ consecutive epochs. 
\end{itemize}

For each training performed, the cutouts in the training set were randomly subdivided into three subsets: $70\%$ for training, $15\%$ for validation and $15\%$ for testing. However, the test subset was not used at this stage.

The structures of the trained networks used the SoftMax activation function for the output layer and the hyperbolic tangent  function (Tanh) for the hidden layers.

\subsubsection{Decision rule}

\hspace{\parindent}All neural networks were designed to classify individual cutouts, and not the epithelium image as a whole.  First, we classify all the cutouts, assigning each one to class $\omega_1$ (dysplastic) or class $\omega_2$ (non-dysplastic). This is done using the following Bayes risk rule:

\begin{equation} \label{eq:DecisionRule}
	\text{Assign $\vx$ to $\omega_1$ ($\omega_2$) if:    } \lambda_{21} P(\omega_1|\vx) >(<) \lambda_{12} P(\omega_2|\vx)
\end{equation}

\noindent in which $\vx$ is the input vector (cutout image); $P(\omega_i|\vx)$ is the posterior conditional probability that $\vx$ is from class $\omega_i$, given the observed cutout image $\vx$ (as estimated by the neural network); and the loss $\lambda_{ij}$ is defined as the loss associated to assigning a cutout to class $\omega_i$ when it actually belongs to class $\omega_j$.

Two different rules were considered to provide diagnosis for each cutout. The first one employs $\lambda_{12} = \lambda_{21}=1$. The second one considers that the loss in having a false negative (erroneously classifying a cutout as non-dysplastic) is twice as large as the loss associated to a false positive. In this case, $\lambda_{12} = 1$ and $\lambda_{21}=2$. The decision rule is then “classify $\vx$ in class $\omega_1$ if”

\begin{equation} \label{eq1}
    {\cfrac{P({\omega}_{1}|\vx)}{P({\omega}_{2}|\vx)} \geq \cfrac{\lambda_{12}}{\lambda_{21}} }
\end{equation}

\noindent The final diagnosis for the case is based on the majority of classifications obtained for all the patient cutouts.

\subsubsection{Performance metrics}

\hspace{\parindent}Different metrics were used to evaluate the performance of the classifiers. The main metric was the confusion matrix, which reports the number of true positives (TP), true negatives (TN), false positives (FP) and false negatives (FN). Using these results we evaluated the sensitivity ($S_e$) and the specificity ($S_p$) of the classifier, given by, respectively

\begin{equation} \label{eq2}
{S_{\rm e}= \cfrac{TP}{TP+FN} }
\end{equation}

\begin{equation} \label{eq3}
{S_{\rm p}= \cfrac{TN}{TN+FP} }
\end{equation}

Another metric employed was the accuracy, evaluated as 

\begin{equation} \label{eq4}
{A_{\rm cc}= \cfrac{TP+TN}{\textit{N}} }
\end{equation}

\noindent where $n$ stands for the total number of data classified. 

\begin{figure}[h]
\centering
\includegraphics[width=0.3\textwidth]{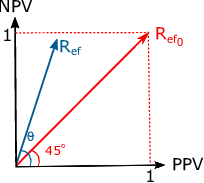}
\caption{Two-dimensional descriptive space of figure of merit $D$. In red: the best classification scenario. In blue: Example of classifier performance.} 
\label{img_figMeritD}
\end{figure}

Finally, we propose in this study a new figure of merit $D$ to evaluate the deviation from an ideal classifier performance. Figure~\ref{img_figMeritD} shows a graphical interpretation of the new figure of merit $D$. The horizontal axis shows the Positive Predictive Value (PPV $=TP$/total number of positives), and the vertical axis shows the Negative Predictive Value (NPV $=TN$/total number of negatives) in the training set. Vector $\bs{R}_{\rm ef}$ represents the composite relative accuracy.  In the ideal classifier (gold standard), $\bs{R}_{\rm ef}=\bs{R}_{{\rm ef}_{0}}$ has magnitude $\sqrt{2}$ and $45^o$ phase. Deviations from this ideal situation represent loss in classification performance. 



To this end, we define $D$ as follows:

\begin{equation} \label{eq5}
    {D= \cfrac{|{D}_{1}|+|{D}_{2}|}{2}}
\end{equation}

\begin{equation} \label{eq6}
    {{D}_{1}= \cfrac{\sqrt{{PPV}^2 + {NPV}^2}}{\sqrt{2}} }
\end{equation}

\begin{equation} \label{eq7}
    {{D}_{2}= \cos{(45^\circ - \theta)}}
\end{equation}

\begin{equation} \label{eq8}
    {\theta = \arctan \frac{NPV}{PPV}}
\end{equation}

In \eqref{eq5}--\eqref{eq8}, ${D}_{1}$ corresponds to the magnitude of vector $\bs{R}_{\rm ef}$, and ${D}_{2}$ corresponds to the angular deviation from the ideal $45^o$ reference. Hence, $0 \le D \le 1$, and the larger the value of $D$ the better the classifier performance. One particularly interesting aspect of the performance criterion $D$ is that it provides an objective way to rank the tested network structures for relative performance. Criterion $D$ is used in the next design phase to estimate the best trained networks.

\subsection{Definition of the best designs}\label{sec:avaDesem}

\hspace{\parindent}At this design stage we had already defined the input dimension (phase 1) and a set of network structures leading to the best performance results, as measured by the average accuracy obtained in design phase 2. The next stage was a statistical performance evaluation, in which we used 100 realizations of training and test for each structure defined in phase 2 with random initialization in search for the best training for each of them. In each realization, the networks were trained with all the cutouts of the 42 cases used in the first and second phases. The testing was performed using cutouts of 31 new cases that had not been used in any previous training or test. The difference in the performances obtained at each of the 100 realizations was due to a different initialization of the network coefficients. For each of the network structures, the coefficients resulting from the training realization that yielded the highest value of criterion $D$ (out of the 100 realizations) for the test set were stored. After these steps, the trained networks, with fixed structures and its coefficients,  were used for the classifications, and the results were compared  with other classifiers.

\subsection{Performance comparisons}

\hspace{\parindent}At this last stage of the project the performances of the networks trained during the classification of the $31$ test cases were compared with the performances in these same cases provided by a convolutional neural network pre-trained by learning transfer, and by trained evaluators (oral pathologists). 

The pre-trained CNN used in our comparison was the Resnet-18 network~ \cite{MathworksResnet}. Some adaptation of the original code was necessary in order to replace the fully connected layer and the sorting layer so that the resulting network output had only two classes as required by the application. In addition, all cutouts were modified to the RGB scale, normalized, and had their sizes readjusted from $256\times256$ to $224\times224$ pixels, to be compatible with the input dimensions of the pre-trained network.

The main CNN training options used were the following:

\begin{itemize}
	\item[a)] Convolutional layer learning rate equal to $10^{-4}$. It was chosen very low so that the filter coefficients and the previously trained weights were not lost.
	\item[b)] Mini-batch size equal to $10$.
	\item[c)] Fully connected layer learning rate equal to 10. It was chosen high so that the learning for these layers was faster than for the convolutional layers. 
	\item[d)] Maximum number of epochs equal to 8. 
\end{itemize}

The training and performance evaluation stages of the CNN network were evaluated using the same methodologies described in sections \ref{sec:2phase} and \ref{sec:avaDesem}.

Finally, to evaluate the usefulness of a classifier in this case, we compared the performances of the best neural classifiers designed with the classifications provided by three trained evaluators. To this end, we asked the trained evaluators to evaluate the original images of each of the $31$ cases used as the test set. The evaluation was made using the whole images (not the cutouts), as would be the case in a real diagnosis situation.

\section{Results}

\hspace{\parindent}In this section we present the performances obtained from the networks designed in the second phase of the project. As mentioned in Section \ref{sec:phase1}, it was determined that the structures with $65,536$ neurons in the input layer had the best performances on average. Hence, all the structures implemented in the second phase had this same input layer size.

\subsection{Definition of the classifier structure}

\hspace{\parindent}Structures with 20, 50, 100, and 150 neurons per layer, and the cost functions mean-square error (MSE) and cross-entropy (CE) were designed. Table~\ref{tab:segFase_semFuncRisc} shows the average classification performances ($50$ realizations) obtained for the $42$ cases available for $\lambda_{12}=\lambda_{21}=1$ in equation \eqref{eq1}. Each realization was characterized by a random initialization of the neural network weights. The majority of structures presented average accuracy higher than $0.80$ with small standard deviations, what is a good indicator of design reliability. Structure 8 presented the highest average accuracy of 0.8205, and its best result reached a 0.8810 accuracy. The results demonstrate that all structures presented high averages for the proposed figure of merit D (equation \eqref{eq5}). However, the best performance was that of structure 8.

\begin{table}[h]
    \renewcommand{\arraystretch}{1.3} 
	\caption{\label{tab:segFase_semFuncRisc}Results - second design phase for $\lambda_{12}=\lambda_{21}=1$.}
    \centering
	\begin{tabular}{lccc}
		
		\hline
		\textbf{Structures} &
		\multicolumn{2}{c}{\textbf{Accuracy}} &
		\multirow{2}{*}{\textbf{\begin{tabular}[c]{@{}c@{}}Figure of \\ merit \footnotemark[5] \\ $\mathbf{D}$($\mathbf{\%}$)\end{tabular}}} \\
		\begin{tabular}[c]{c@{}@{}}(NHL\footnotemark[1]/NNPHL\footnotemark[2]/\\ Cost function)\end{tabular} &
		\textbf{Average$\pm$SD\footnotemark[3]} &
		\textbf{\begin{tabular}[c]{@{}@{}c@{}}Max.\\ value \footnotemark[4]\end{tabular}} &
		\\ \hline
		\textbf{1}~(2/20/MSE) & 0.7881$\pm$0.0334 & 0.8571 & 89.25  \\
		\textbf{2}~(2/20/CE\footnotemark[6])  & 0.8019$\pm$0.0386 & 0.8810  & 89.93  \\
		\textbf{3}~(2/50/MSE) & 0.8014$\pm$0.0421 & 0.9048  & 89.94  \\
		\textbf{4}~(2/50/CE\footnotemark[6])  & 0.8143$\pm$0.0340 & 0.8810 & 90.58  \\
		\textbf{5}~(2/100/MSE) & 0.8105$\pm$0.0391 & 0.8810  & 90.36  \\
		\textbf{6}~(2/100/CE\footnotemark[6]) & 0.8086$\pm$0.0376 & 0.8810  & 90.26  \\
		\textbf{7}~(2/150/MSE)  & 0.8043$\pm$0.0507 & 0.9048 & 90.05  \\
		\textbf{8}~(2/150/CE\footnotemark[6]) & 0.8205$\pm$0.0382 & 0.8810 & 90.88 \\ \hline
	\end{tabular}
\end{table}
\hspace{\parindent} \hspace{\parindent} \hspace{\parindent} \footnotemark[1]{NHL - Number of Hidden Layers.}

\hspace{\parindent} \hspace{\parindent} \hspace{\parindent} \footnotemark[2]{NNPHL - Number of Neurons Per Hidden Layer.}

\hspace{\parindent}  \hspace{\parindent} \hspace{\parindent} \footnotemark[3]{SD - Standard Deviation.}
 
\hspace{\parindent}  \hspace{\parindent} \hspace{\parindent} \footnotemark[4]{Max. value - Maximum value.}

\hspace{\parindent}  \hspace{\parindent} \hspace{\parindent} \footnotemark[5]{Percentage of maximum possible value in \eqref{eq5}.}

\hspace{\parindent}  \hspace{\parindent} \hspace{\parindent} \footnotemark[6]{CE - Cross-Entropy.} \\

\begin{figure}[h!]
\centering
\includegraphics[width=0.5\textwidth]{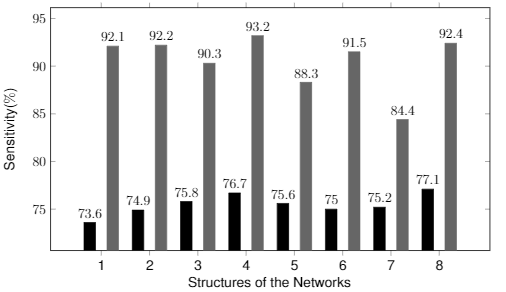}
\caption{Average sensitivities of network structures with two hidden layers. Black: $\lambda_{12}=\lambda_{21}=1$. Gray: $\lambda_{12}=1$, $\lambda_{21}=2$.} 
\label{img_sensi2fase}
\end{figure}

\begin{figure}[h!]
\centering
\includegraphics[width=0.5\textwidth]{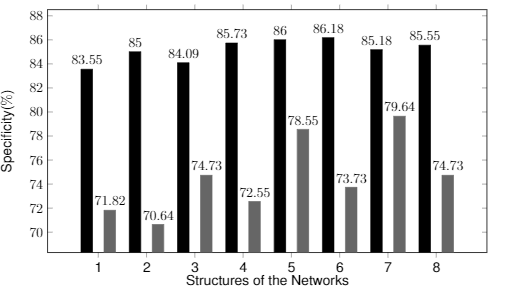}
\caption{Average specificities of network structures with two hidden layers. Black: $\lambda_{12}=\lambda_{21}=1$. Gray: $\lambda_{12}=1$, $\lambda_{21}=2$.} 
\label{img_espec2fase}
\end{figure}

The same eight structures in Table~\ref{tab:segFase_semFuncRisc} were trained using $\lambda_{12}=1$ and $\lambda_{21}=2$ in~\eqref{eq1} to test the design performance with a conservative risk function. It was observed that the average accuracy and mean values of D did not present a significant increase when compared to the values in Table~\ref{tab:segFase_semFuncRisc}. Figures \ref{img_sensi2fase} and \ref{img_espec2fase} show, respectively, comparisons of the sensitivity and specificity values obtained using the two different risk functions. As expected, the sensitivity increased and the specificity reduced when $\lambda_{21}=2$ was used. This is because, in this case, a greater importance was given to the occurrence of false negatives (erroneous classification of dysplastic epithelia). As a consequence, a greater number of cases were classified by the network as dysplastic, increasing the number of true positives and the average sensitivity. On the other hand, the number of false positives increased, decreasing the average specificity.
 
For the next design step we selected the three structures in Table~\ref{tab:segFase_semFuncRisc} which yielded the largest average accuracies, namely structures 4, 5, and 8. For these three structures we proceeded to increase the number of hidden layers while such increase resulted in clear possibility of performance improvement.

Table \ref{tab:struct4} shows that the performance of structure 4 degraded when a third hidden layer was added. Table \ref{tab:struct5} shows the performance of structure $5$. It is noted that a performance increase was obtained with $4$ hidden layers (structure $11$). Table \ref{tab:struct8} shows the performance of structure $8$. In this case, an improved performance was verified as the number of hidden layers increased to 3 (structure $13$). Comparing the results for structures $4$, $11$ and $13$, structure $13$ was the best one, with an average accuracy of 0.8271, and an average figure of merit D above $91\%$. Given their comparable performances, we considered these three structures (4, 11, and 13) in the following design and comparison stages.

\begin{table}[h!]
    \renewcommand{\arraystretch}{1.3} 
	\caption{\label{tab:struct4}Performance of structure $4$ with 2 and 3 hidden layers ($\lambda_{12}=\lambda_{21}=1$).}
	\centering
	\resizebox{9cm}{!}{
		\begin{tabular}{lccllc}
			\hline
			\textbf{Structures} &
			\multicolumn{2}{c}{\textbf{Accuracy}} &
			\multicolumn{1}{c}{$\mathbf{S_{\rm e}}$\footnotemark[1]} &
			\multicolumn{1}{c}{$\mathbf{S_{\rm p}}$\footnotemark[2]} &
			\multirow{2}{*}{\textbf{\begin{tabular}[c]{@{}c@{}}Figure \\ of merit  \\ $D$ ($\%$)\end{tabular}}} \\
			\begin{tabular}[c]{@{}c@{}}(NHL/NNPHL\\ Cost function)\end{tabular} &
			\textbf{Average$\pm$SD} &
			\textbf{\begin{tabular}[c]{@{}c@{}}Max.\\ value \end{tabular}} &
			\multicolumn{1}{c}{($\%$)} &
			\multicolumn{1}{c}{($\%$)} &
			\\ \hline
			\textbf{4}~(2/50/CE)  & 0.8143$\pm$0.0340 & 0.8810 & 76.70 & 85.73 & 90.58  \\
			\textbf{9}~(3/50/CE)  & 0.8105$\pm$0.0363 & 0.9048 & 76.10 & 85.55 & 90.38   \\ \hline
		\end{tabular}
	}
\end{table}
\hspace{\parindent} \hspace{\parindent} \hspace{\parindent} \hspace{\parindent} \footnotemark[1]{$S_{\rm e}$ - Sensitivity.}

\hspace{\parindent} \hspace{\parindent} \hspace{\parindent} \hspace{\parindent} \footnotemark[2]{$S_{\rm p}$ - Specificity.} \\

\begin{table}[h!]
    \renewcommand{\arraystretch}{1.3} 
	\caption{\label{tab:struct5}Performance of structure $5$ with $2-5$ hidden layers ($\lambda_{12}=\lambda_{21}=1$).}
	\centering
	\resizebox{9cm}{!}{
		\begin{tabular}{lccllc}
			\hline
			\textbf{Structures} &
			\multicolumn{2}{c}{\textbf{Accuracy}} &
			\multicolumn{1}{c}{$\mathbf{S_{\rm e}}$} &
			\multicolumn{1}{c}{$\mathbf{S_{\rm p}}$} &
			\multirow{2}{*}{\textbf{\begin{tabular}[c]{@{}c@{}}Figure \\ of merit  \\ $D$ ($\%$)\end{tabular}}} \\
			\begin{tabular}[c]{@{}c@{}}(NHL/NNPHL/\\ Cost function)\end{tabular} &
			\textbf{Average$\pm$SD} &
			\textbf{\begin{tabular}[c]{@{}c@{}}Max.\\ value \end{tabular}} &
			\multicolumn{1}{c}{($\%$)} &
			\multicolumn{1}{c}{($\%$)} &
			\\ \hline
			\textbf{5}~(2/100/MSE)  & 0.8105$\pm$0.0391 & 0.8810 & 75.60 & 86.00 & 90.36  \\
			\textbf{10}~(3/100/MSE)  & 0.8100$\pm$0.0515 & 0.8810 & 77.20 & 84.45 & 90.39  \\
			\textbf{11}~(4/100/MSE)  & 0.8195$\pm$0.0411 & 0.9048 & 78.80 & 84.82 & 90.88  \\
			\textbf{12}~(5/100/MSE)  & 0.8071$\pm$0.0394 & 0.8810 & 77.20 & 83.91 & 90.25  \\ \hline
		\end{tabular}
	}

\end{table}

The performances of structures $4$, $5$ and $8$ obtained with the increase of the hidden layers for $\lambda_{12}=1$ and $\lambda_{21}=2$ were also verified. The relative performances were very similar to those shown in Figures \ref{img_sensi2fase} and \ref{img_espec2fase}, namely, the average sensitivities increased while the average specificities reduced. The average accuracies and figure of merit D had no significant changes when compared with the results obtained from the classifications using $\lambda_{12}=\lambda_{21}=1$ (Tables \ref{tab:struct4}, \ref{tab:struct5} and \ref{tab:struct8}). Hence, the results are presented only for the latter case.

\begin{table}[h!]
    \renewcommand{\arraystretch}{1.3} 
	\caption{\label{tab:struct8}Performance of structure $8$ with $2-4$ hidden layers ($\lambda_{12}=\lambda_{21}=1$).}
	\centering
	\resizebox{9cm}{!}{
		\begin{tabular}{lccllc}
			\hline
			\textbf{Structures} &
			\multicolumn{2}{c}{\textbf{Accuracy}} &
			\multicolumn{1}{c}{$\mathbf{S_{\rm e}}$} &
			\multicolumn{1}{c}{$\mathbf{S_{\rm p}}$} &
			\multirow{2}{*}{\textbf{\begin{tabular}[c]{@{}c@{}}Figure \\ of merit  \\ $D$ ($\%$)\end{tabular}}} \\
			\begin{tabular}[c]{@{}c@{}}(NHL/NNPHL/\\ Cost function)\end{tabular} &
			\textbf{Average$\pm$SD} &
			\textbf{\begin{tabular}[c]{@{}c@{}}Max.\\ value \end{tabular}} &
			\multicolumn{1}{c}{($\%$)} &
			\multicolumn{1}{c}{($\%$)} &
			\\ \hline
			\textbf{8}~(2/150/CE)  & 0.8205$\pm$0.0382 & 0.8810 & 77.10 & 86.55 & 90.88  \\
			\textbf{13}~(3/150/CE)  & 0.8271$\pm$0.0343 & 0.8571 & 78.80 & 86.27 & 91.25   \\
			\textbf{14}~(4/150/CE)  & 0.8110$\pm$0.0313 & 0.8571 & 75.40 & 86.27 & 90.38   \\ \hline
		\end{tabular}
	}

\end{table}

\subsection{Performance evaluation stage}

\hspace{\parindent}The figure of merit D was computed for the test cases using \eqref{eq5} after each training. The trained networks with each structure yielding the highest value of D had their coefficients stored to be used in future comparisons.  Table \ref{tab:bestsStruc} shows the performance of the best design for each of the three selected structures.

\begin{table}[h!]
    \renewcommand{\arraystretch}{1.3} 
	\caption{\label{tab:bestsStruc}Performance of the three best structures ($\lambda_{12}=\lambda_{21}=1$).}
	\centering
	\begin{tabular}{lcccc}
		\hline
		\textbf{Structures} & \textbf{Accuracy} & \textbf{$\mathbf{S_{\rm e}}$} & \textbf{$\mathbf{S_{\rm p}}$} & \multirow{2}{*}{\textbf{\begin{tabular}[c]{@{}c@{}}Figure \\ of merit \\ $D$ ($\%$)\end{tabular}}} \\
		\begin{tabular}[c]{@{}c@{}}(NHL/NNPHL/\\ Cost function)\end{tabular} & \textbf{} & ($\%$) & ($\%$) &       \\ \hline
		\textbf{4}~(2/50/CE)                & 0.8065    & 81.25  & 80     & 90.31 \\
		\textbf{11}~(4/100/MSE)             & 0.8710    & 81.25  & 93.33  & 93.63 \\
		\textbf{13}~(3/150/CE)              & 0.8710    & 81.25  & 93.33  & 93.63 \\ \hline
	\end{tabular}
	
\end{table}

Table \ref{tab:CNN_train} shows the average performances of all four classifiers in the training stage. The pre-trained CNN yielded a slightly better average accuracy and the best performance when compared to the MLP networks. However, its accuracy had the largest standard deviation (about 10\% higher than structure 11, and 30\% higher than the best MLP structure 13).

\begin{table}[h!]
    \renewcommand{\arraystretch}{1.3} 
	\caption{\label{tab:CNN_train}Performance comparison with pre-trained CNN - training.}
	\centering
	\resizebox{9cm}{!}{
		\begin{tabular}{lccllc}
			\hline
			\textbf{Structures} &
			\multicolumn{2}{c}{\textbf{Accuracy}} &
			\multicolumn{1}{c}{$\mathbf{S_{\rm e}}$} &
			\multicolumn{1}{c}{$\mathbf{S_{\rm p}}$} &
			\multirow{2}{*}{\textbf{\begin{tabular}[c]{@{}c@{}}Figure \\ of merit  \\ $D$ ($\%$)\end{tabular}}} \\
			\begin{tabular}[c]{@{}c@{}}(NHL/NNPHL/\\ Cost function)\end{tabular} &
			\textbf{Average$\pm$SD} &
			\textbf{\begin{tabular}[c]{@{}c@{}}Max.\\ value \end{tabular}} &
			\multicolumn{1}{c}{($\%$)} &
			\multicolumn{1}{c}{($\%$)} &
			\\ \hline
			\textbf{4}~(2/50/CE)  & 0.8143$\pm$0.0340 & 0.8810 & 76.70 & 85.73 & 90.58  \\
			\textbf{11}~(4/100/MSE)  & 0.8195$\pm$0.0411 & 0.9048 & 78.80 & 84.82 & 90.88   \\ 
			\textbf{13}~(3/150/CE)  & 0.8271$\pm$0.0343 & 0.8571 & 78.80 & 86.27 & 91.25   \\ 
			\textbf{Pre-trained}~\textbf{CNN}  & 0.8505$\pm$0.0449 & 0.9524 & 83.30 & 86.64 & 92.47   \\ \hline
		\end{tabular}
	}

\end{table}

Table \ref{tab:CNN_avaDesem} shows the obtained classification results for the test set. The best performance was obtained from the CNN. 

\begin{table}[h!]
    \renewcommand{\arraystretch}{1.3} 
	\caption{\label{tab:CNN_avaDesem} Performance comparison with pre-trained CNN - test.}
	\centering
	\begin{tabular}{lcccc}
		\hline
		\textbf{Structures} & \textbf{Accuracy} & \textbf{$\mathbf{S_{\rm e}}$} & \textbf{$\mathbf{S_{\rm p}}$} & \multirow{2}{*}{\textbf{\begin{tabular}[c]{@{}c@{}}Figure \\ of merit \\ $D$ ($\%$)\end{tabular}}} \\
		\begin{tabular}[c]{@{}c@{}}(NHL/NNPHL/\\ Cost function)\end{tabular} & \textbf{} & ($\%$) & ($\%$) &       \\ \hline
		\textbf{4}~(2/50/CE)                & 0.8065    & 81.25  & 80     & 90.31 \\
		\textbf{11}~(4/100/MSE)             & 0.8710    & 81.25  & 93.33  & 93.63 \\
		\textbf{13}~(3/150/CE)              & 0.8710    & 81.25  & 93.33  & 93.63 \\
		\textbf{pre-trained}~\textbf{CNN} & 0.9032    & 93.75  & 86.67  & 95.10 \\\hline
	\end{tabular}
	
\end{table}

\begin{figure}[h!]
\centering
\includegraphics[width=0.4\textwidth]{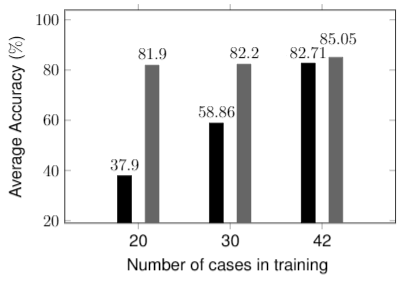}
\caption{Average accuracy of classifiers with the increase in the number of cases used in training. Black: structure $13$. Gray: pre-trained CNN.} 
\label{img_aumentoCasos_train}
\end{figure}

As an additional evaluation, it is of interest to gauge the potential of the different structures as more samples are available for training. To this end, we have trained the classifiers using both the CNN and structure $13$ for an increasing number of training samples. Figure \ref{img_aumentoCasos_train} shows the progress in average accuracy obtained by the two classifiers when training is performed with 20, 30, and 42 cases. These results indicate that the performance of classifiers using MLP networks tends to approach that of the CNN classifier as the amount of available training data increases.  

\subsection{Comparison with trained evaluators}

\hspace{\parindent}Table \ref{tab:compAvalia} compares the performances of the neural classifiers and the three evaluators in the classification of the $31$ cases of the test set. This table shows that neural classifiers generated an accuracy equal to or even higher than the three trained evaluators. The three evaluators correctly  classified all dysplastic cases (sensitivities were $100\%$). However, they presented specificities equal or inferior to the neural classifiers.

\begin{table}[h!]
    \renewcommand{\arraystretch}{1.3} 
	\caption{\label{tab:compAvalia} Comparison including trained evaluators - test set.}
	\centering
	\begin{tabular}{lcccc}
		\hline
		\textbf{Structures} & \textbf{Accuracy} & \textbf{$\mathbf{S_{\rm e}}$} & \textbf{$\mathbf{S_{\rm p}}$} & \multirow{2}{*}{\textbf{\begin{tabular}[c]{@{}c@{}}Figure \\ of merit \\ $D$ ($\%$)\end{tabular}}} \\
		\begin{tabular}[c]{@{}c@{}}(NHL/NNPHL/\\ Cost function)/\\ \textbf{Evaluator}\end{tabular} & \textbf{} & ($\%$) & ($\%$) &       \\ \hline
		\textbf{4}~(2/50/CE)                & 0.8065    & 81.25  & 80     & 90.31 \\
		\textbf{11}~(4/100/MSE)             & 0.8710    & 81.25  & 93.33  & 93.63 \\
		\textbf{13}~(3/150/CE)              & 0.8710    & 81.25  & 93.33  & 93.63 \\
		\textbf{pre-trained}~\textbf{CNN}  & 0.9032    & 93.75  & 86.67  & 95.10 \\
		\textbf{Evaluator 1} & 0.9032    & 100  & 80  & 94.97 \\
		\textbf{Evaluator 2} & 0.6451    & 100  & 26.67  & 79.86 \\
		\textbf{Evaluator 3} & 0.8710    & 100  & 73.34 & 93.26 \\ \hline
	\end{tabular}
\end{table}

\section{Discussion}

\hspace{\parindent}The diagnosis of epithelial dysplasia is provided by oral pathologists based on visual analysis of histopathological images. However, this diagnosis procedure may be subjective and dependent on the professional experience of the pathologist. In this work, a diagnostic aid system was proposed to help to reduce this subjectivity and the variability in diagnoses. The objective was to develop an epithelial classifier with low computational cost, using the expertise of pathologists to avoid excessive complexity. A methodology was proposed for the design of a simple MLP neural network classifier. Two of the resulting classifiers (11 and 13 in Table \ref{tab:bestsStruc}) yielded an average accuracy of $87\%$ in the performance assessment stage. 

It has been recognized in the technical literature that Convolutional Neural Networks (CNNs) tend to yield excellent performance in image classification, especially in the medical field \cite{art2}. Hence, CNNs qualify as the best candidates for performance comparison with the oral lesion classifiers designed using the proposed approach. However, training a CNN from scratch requires a huge amount of data, in the order of millions of samples. Such amount of data is hardly available in the medical area, and such a solution (if ever feasible) would result in a time-consuming and very costly training process. A typical solution to circumvent this limitation is to apply transfer learning. In this method, a fine-tuning is done in a pre-trained network, where the weights of the initial layers are frozen to values obtained from training with general images, and only the last layers are trained with the application-specific data \cite{CNNNove}. This was the possible solution in our case, due to the limited amount of data available (a typical situation for the application at hand).

The performances of the best classifier designs were compared with that of a pre-trained CNN, the state-of-the-art in image classifiers for medical applications. The results (Table \ref{tab:CNN_avaDesem}) show an accuracy of more than 87\%, only 3.7\% smaller than that obtained from the CNN solution. The sensitivity of the CNN was 15\% better (93.75\% versus 81.25\%), but at the cost of a $7\%$ worse specificity (86.67\% versus 93.33\% of the proposed solution). The difference in the value of the newly proposed figure of merit was only 1.6\% in favor fo the CNN solution (95.1\% versus 93.63\%). It should be noted that this good performance of the proposed solution, which is quite comparable to that of the pre-trained CNN, comes at a significant advantage in complexity. 


Table \ref{tab:compComplexComp} compares the computational cost in Floating-point Operations (FLOPs) of the three classifiers designed using the proposed methodology with that of the pre-trained CNN. It is noted that the pre-trained CNN has an operation complexity at least $100$ times greater than the most complex network among networks $4$, $11$ and $13$. This increase in complexity is also accompanied by a significant increase in the amount of memory required for the CNN, when compared to the other networks. This much higher complexity does not justify the corresponding modest increase in performance, showing that the contribution of theoretical expertise to the design of the classifier can easily surpass the advantages of using very sophisticated neural network structures to classify raw data. 

\begin{table}[h!]
    \renewcommand{\arraystretch}{1.3} 
	\centering
	\caption{\label{tab:compComplexComp}Comparison of computational cost.}

	\begin{tabular}{lc}
		\hline
		\textbf{Structures} & \textbf{\begin{tabular}[c]{@{}c@{}} Computational \\ cost\\ (FLOPs)\end{tabular}} \\ \hline
		\textbf{4}~(2/50/Cross-Entropy) & 6.58 M \\
		\textbf{11}~(4/100/MSE) & 13.20 M \\
		\textbf{13}~(3/150/Cross-Entropy) & 19.79 M \\
		\textbf{pre-trained}~\textbf{CNN} ResNet-18 & $\cong$ 2 G \\ \hline
	\end{tabular}
\end{table}

We have also verified the performance of the classifiers when using different number of cases during their training. Our results (Figure \ref{img_aumentoCasos_train}) showed that the performance of the proposed MLP classifiers converges to that of the pre-trained CNN as more cases are used for training. Hence, weighting implementation cost and classification performance, the generally accepted superiority of a CNN solution for any image classification application is clearly open to question.

Another important aspect of the specific application is to address the value of the proposed classifiers as a supporting tool for the pathologist in reaching the correct diagnosis. The value of such support should be evaluated considering the fact that a diagnosis of epithelial dysplasia is rarely made by a large number of experts. Also, even a group of highly trained professionals may reach diverse conclusions, especially in less obvious cases. Mathematically, these two aspects indicate a tendency of large variability in the diagnosis made by a small set of pathologists. Recognizing such a tendency, it is of interest of the pathologist to have the support of a well designed classifier when analyzing oral lesions for detecting dysplasia.

This tendency of high variability among trained evaluators was observed in Table~\ref{tab:compAvalia}, where average accuracy of evaluators was $0.8064$, but with standard deviation of $0.1406$, while the average accuracy of the three MLP networks was $0.8495$, with standard deviation of $0.0372$. Although these results are not statistically significant due to the small number of trained evaluators, they correspond to a practical situation, as the diagnosis will rarely be made by considering a large number of pathologists opinions. These results corroborate the expected tendency of having a large variability in classifications by few trained evaluators, which suggests that the help of a well trained algorithmic classifier should be welcome as a support for decision.

Another interesting observation in Table~\ref{tab:compAvalia} is that the three evaluators correctly classified all dysplastic cases (sensitivities were $100\%$). However, the trained evaluators yielded specificities equal to or lower than those obtained with neural classifiers. This fact suggests that the trained evaluators tended to classify cases as dysplastic when there was doubt in the classification. This tendency to be "on the safe side" leads to and increase in the number of false positives, which corresponds to using a risk function with $\lambda_{21}>1$ for reaching the decision.

\section{Conclusions}\label{sec:conclusions}

\hspace{\parindent}This study showed that the multilayered network structures combined with the pathologists' knowledge to choose the cutout region that would be delivered to the network, presented performances similar to those of the network that is considered the state of the art in image classifications (pre-trained CNN), with considerable less operational complexity. In addition, it was analyzed that increasing the number of cases used in the training of MLP networks would bring this performances even closer.

Finally, the average performance of three trained evaluators was compared with the average performance of the three MLP networks. We observed that they resulted in very close averages of accuracy, but the standard deviation of neural structures was approximately $74\%$ lower than the standard deviation of trained evaluators. This high variability in the diagnoses of trained evaluators may be associated with their emotional state during the classification of cases. Thus, using a well trained classifier to aid in the diagnosis, could be welcome to reduce this high variability found.
    
\section{Acknowledgments}   

\hspace{\parindent}This study was supported in part by the Coordenação de Aperfeiçoamento de Pessoal de Nível Superior – Brasil (CAPES) – Finance Code 001, and by the National Council for Scientific and Technological Development (CNPq).

\bibliographystyle{IEEEbib}
\bibliography{bibliography}

\end{document}